\documentclass[prl,twocolumn,amsmath,amssymb,showpacs,showkeys,superscriptaddress]{revtex4}
\usepackage{graphicx}
\usepackage{dcolumn}
\usepackage{hyperref}
\newcommand{\be}{\begin{equation}}
\newcommand{\ee}{\end{equation}}
\newcommand{\ba}{\begin{eqnarray}}
\newcommand{\ea}{\end{eqnarray}}
\begin{document}
\title{Entropy-based measure of structural order in water}
\author{Rubens Esposito}
\email{\tt rubens.re@libero.it} \affiliation{Universit\`a degli Studi di
Messina,\\Dipartimento di Fisica, Contrada Papardo, 98166 Messina,
Italy}
\author{Franz Saija}
\email{\tt saija@me.cnr.it}
\affiliation{CNR - Istituto per i Processi Chimico-Fisici,
Sezione di Messina,\\Via La Farina 237, 98123 Messina, Italy}
\author{A. Marco Saitta}
\email{\tt marco.saitta@impmc.jussieu.fr} \affiliation{D\'epartement
de Physique des Milieux Denses, IMPMC, CNRS-UMR 7590, B115,
Universit\'e Pierre et Marie Curie, F-75252 Paris, France}
\author{Paolo V. Giaquinta}
\altaffiliation{Corresponding author} \email{\tt
paolo.giaquinta@unime.it} \affiliation{Universit\`a degli Studi di
Messina,\\Dipartimento di Fisica, Contrada Papardo, 98166 Messina,
Italy}

\date{\today}

\begin{abstract}
We analyze the nature of the structural order established in liquid TIP4P water in the framework provided by the
multi-particle correlation expansion of the statistical entropy. Different regimes are mapped onto the phase diagram of
the model upon resolving the pair entropy into its translational and orientational components. These parameters are used
to quantify the relative amounts of positional and angular order in a given thermodynamic state, thus allowing a
structurally unbiased definition of low-density and high-density water. As a result, the structurally anomalous region
within which both types of order are simultaneously disrupted by an increase of pressure at constant temperature is
clearly identified through extensive molecular-dynamics simulations.

\end{abstract}

\pacs{05.20.Jj, 61.20.Ja, 64.10.+h, 64.70.Ja}

\keywords{water, phase diagram, structural order, TIP4P, entropy, pair correlations}

\maketitle
Water undergoes, under compression, a continuous structural transformation from an open, hydrogen-bonded tetrahedral
structure to a denser form in which the local molecular arrangement beyond the first coordination shell looks deeply
modified ~\cite{soper2000}. Such two forms are ordinarily referred to as low-density and high-density water (LDW, HDW).
There is wide consensus on their being high-temperature manifestations of two glassy polymorphs of ice: low-density
and high-density amorphous ice, respectively~\cite{mishima1998}. Raman and Brillouin spectroscopic studies
have recently given conflicting predictions on the thermodynamic boundary between LDW and HDW
forms~\cite{kawamoto2004,li2005}. In this Letter we present an approach to analyze different ordering regimes in water that is entirely based on
entropy and which can be used to trace the LDW-HDW boundary without resorting to any {\it a
priori} assumptions on the local structure of the stable liquid phase.

Quantifying the degree of structural order present in liquids and amorphous materials is a challenging issue that has
received considerable attention in the last few years~\cite{torquato2000,truskett2000,kansal2002,errington2003}. The
possibility of characterizing in a synthetic way the macroscopic
state of a non-crystalline substance through some integrated measure
of the overall amount of order emerging from microscopic
correlations has been promisingly exploited also for complex
molecular liquids such as water~\cite{errington2001,lynden-bell2005,giovambattista2005}. The
method is essentially based on the identification of appropriate
metrics for both positional and angular order. The thermodynamic
states of the system can then be mapped onto a structural diagram,
spanned by such two parameters, which can be used to obtain a direct
and quantitative insight into different ordering regimes achieved by
materials under cooling and/or compression. In this Letter we
introduce and analyze a metric for discussing structural anomalies
in water that is entirely and consistently based, for both
translational and bond-orientational order, on the ``fine
structure'' of the configurational entropy resolved through spatial
correlations of increasing order.

The basic relation between entropy and order can be cast in a
formally rigorous statistical-mechanical framework by resorting to
the multi-particle correlation expansion of the configurational
entropy~\cite{green1952,ng}. For a classical, homogeneous and
isotropic fluid, this expansion has the form: \ba\label{eq:sex}
s_{\rm ex} = \sum_{n=2}^{\infty} s_{n}\ , \ea \noindent where
$s_{\rm ex}$ is the excess entropy per particle in units of the
Boltzmann constant and the partial entropies $s_{n}$ are obtained
from the integrated contributions of spatial correlations between
{\it n}-tuples of particles. The pair entropy of a molecular fluid
can be written as \cite{lapa,laka}: \be \label{eq:s2mol} s_{2} =
-\frac{1}{2} \frac{\rho}{\Omega^2} \int \{ g({\bf r},\omega^2) \ln
[g({\bf r},\omega^2)] - g({\bf r},\omega^2) + 1\} {\rm d} {\bf r}
{\rm d} \omega^2 \,, \ee \noindent where $\rho$ is the number
density and $g({\bf r},\omega^2)$ is the pair distribution function
(PDF) which depends on the relative separation $\bf r$ between a
pair of molecules and on the set of Euler angles
$\omega^2\equiv[\omega_{1},\omega_{2}]$ that specify the relative
orientations of the two particles in the laboratory reference frame.
The quantity $\Omega$ represents the integral over the Euler angles
of one molecule ($\Omega=8\pi^2$ for non-linear molecules). The pair
entropy typically accounts for the overwhelming contribution to the
total excess entropy of a dense liquid. In order to highlight the
relative contributions of translational and orientational
correlations \be \label{s:s2} s_{2}= s_{2}^{\rm (tr)} + s_{2}^{\rm
(or)} \,, \ee we follow Lazaridis and Karplus~\cite{laka} and
factorize the PDF as \be \label{s:pdf} g(r,\omega^{2}) =
g(r)g(\omega^{2}|r) \,. \ee \noindent In Eq.~\ref{s:pdf}, $g(r)$ is
the radial distribution function (RDF) for some arbitrary site in
the molecules which, in the case of water, is identified with the
oxygen atom. This function is obtained upon integration of the PDF
over its angular coordinates. Instead, the function
$g(\omega^{2}|r)$ represents the conditional distribution function
for the relative orientation of a pair of molecules whose relative
distance is $r$. This quantity is normalized to $\Omega^{2}$ and is
referred to as the orientational distribution function (ODF). Using
Eq.~\ref{s:pdf}, one obtains for the translational and orientational
contributions to the pair entropy: \be \label{s:s2gl} s_{2}^{\rm
(tr)} = -\frac{1}{2} \rho \int [ g(r) \ln g(r) - g(r) + 1 ] {\rm d}
{\bf r} \,, \ee and \be \label{s:s2or} s_{2}^{\rm (or)} = \rho \int
g(r)S(r) {\rm d} {\bf r}\,, \ee with \be \label{s:s2orbig} S(r) =
-\frac{1}{2}\frac{1}{\Omega^2} \int g(\omega^{2}|r) \ln
[g(\omega^{2}|r)] {\rm d} \omega^{2} \;. \ee The use of the
positive-definite quantity $-s_{2}^{\rm (tr)}$ as a measure of
translational order in both atomic and molecular fluids has been
already suggested by Truskett and coworkers~\cite{truskett2000} and
by Errington and Debenedetti~\cite{errington2001} as an alternative
to the integral of the absolute value of the total correlation
function, $\bigl |{g(r)-1} \bigr |$, over a finite number of
coordination shells. As requested for an order parameter, both
quantities vanish in an ideal gas. As for the orientational order,
the metric exploited so far for water is actually a measure of the
degree of tetrahedrality in the average distribution of the four
oxygen atoms that are closest to a given central
atom~\cite{chau1998,errington2001,lynden-bell2005,giovambattista2005}. As
such, the definition of this order parameter rests on the knowledge
of a pre-defined local structure. We propose, instead, to adopt
$-s_{2}^{\rm (or)}$ as an independent measure of angular order, {\it
au pair} with $-s_{2}^{\rm (tr)}$ for positional order. This
mutually consistent choice for the translational and orientational
metrics is rooted on a general and rigorous statistical-mechanical
ground. As such, it can be implemented for any molecular fluid,
without any {\it a priori} cognition of the local environment.

We calculated the translational and orientational pair entropies for
the TIP4P model of liquid water~\cite{Jorgensen83}. Although more
and more sophisticated classical potentials are still being
developed~\cite{jorge2,Mahoney00,Guillot01,Nada03,Rousseau,abascal2005},
the TIP4P description is, on the overall, very satisfactory when
compared with experiments~\cite{Sorenson00,Saitta04}, and its
predictive power is well established~\cite{Mishima84,Saitta03}.

We carried out molecular-dynamics (MD) calculations, implemented through the PINY code~\cite{piny} on a system of $108$
water molecules in a cubic supercell with periodic boundary conditions. The sensitivity of the results to the size of the
sample was tested with a number of simulations performed at ambient pressure on a system of 512 molecules. Long-range
forces were computed through a particle-mesh Ewald method. The distance cutoff was set at $8.5 \, \rm \AA$. The TIP4P
model is based on a rigid-molecule description, in that it neglects intramolecular degrees of freedom. A time step of
$2.5 \, \rm fs$ turns out to be sufficient to ensure a proper dynamical evolution. Simulation times were in the $2.0
\, \rm ns$ range. The configurations were stored every $1-2 \, \rm ps$. The resulting RDFs were calculated over
1000-2000 different configurations with a spatial resolution $\Delta r=0.05 \rm \, {\AA}$. The ODF was calculated at
intervals of $10^{\circ}$. We performed independent simulations in the NPT ensemble for pressures comprised between
ambient pressure and $10 \, \rm kbar$ in the $210-400 \, \rm K$ temperature range, with a step of $10 \, \rm K$.
Lower-density states ($0.95, 0.90, 0.85, 0.80, 0.75 \, \rm g/cm^3$), corresponding to negative pressures, were
investigated with constant-volume simulations carried out over the same temperature range. The initial set of positions
and velocities of each run was taken from the last configuration of a simulation carried out at the same pressure, at
an immediately higher or lower temperature.

At variance with $s_{2}^{\rm (tr)}$, the calculation of $s_{2}^{\rm
(or)}$ poses severe difficulties. Indeed, computing and integrating
the ODF - which, in principle, depends on up to nine independent
variables - directly for a polyatomic molecular fluid over an
extended range of temperatures and pressures is a formidable task.
In practice, in the case of water, the ODF is a function of six
variables only: \emph{i.e.}, the intermolecular distance, the two
angles formed by the dipole vector of each molecule with the
intermolecular axis, the angle describing the rotation of the two
molecules around the intermolecular axis, and the two angles
describing the rotation of each water molecule around its dipole
vector. In order to tackle this fairly demanding computational task,
we resorted to an approximate mapping of the ODF onto its
low-density form. This mapping, known as ``adjusted gas-phase" (AGP)
approximation, was introduced by Lazaridis and Karplus~\cite{laka}
and is implemented under a two-fold constraint: i) the requirement
that the ``exact'' - {\it i.e.}, numerically evaluated -
orientationally averaged interaction energy between two water
molecules be reproduced, as a function of their relative distance,
in the {\it liquid} phase; ii) the consistency of the approximate
ODF with the one- and two-dimensional projections of the exact
function over the five-dimensional space spanned by the angular
coordinates. The reliability of the AGP approximation has been
positively checked through a number of indicators such as the
oxygen-oxygen and oxygen-hydrogen site-site distribution functions.
An independent direct calculation~\cite{zielkiewicz2005} of the
orientational pair entropy of 512 TIP4P water molecules carried out
at constant volume ($\rho = 0.999 \, \rm g/cm^3$) at $T=298 \, \rm
K$ leads to a value ($-s_{2}^{\rm (or)}=50.08 \, \rm Joule/mol \,
K$,~\cite{zielkiewicz2005bis}) that is in acceptable agreement with
the constant-pressure (1 atm) AGP estimates obtained with 216
molecules at the same temperature ($-s_{2}^{\rm (or)}=48.74 \, \rm
Joule/mol \, K$,~\cite{lazaridis}), and with 512 molecules at $T=300 \,
\rm K$ ($-s_{2}^{\rm (or)}=48.70 \, \rm Joule/mol \, K$, this work).
%
%
\begin{figure}
\includegraphics[width=8cm,angle=0]{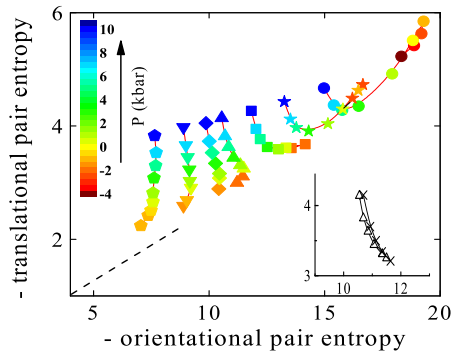}
\caption{\label{fig1} (Color online) Ordering map: Translational order parameter,
$-s_{2}^{\rm (tr)}$, plotted as a function of the orientational
order parameter, $-s_{2}^{\rm (or)}$, for $108$ particles at
different temperatures (entropy units: cal/mol K). Circles, $T=210
\, \rm K$; stars, $T=240 \, \rm K$; squares, $T=270 \, \rm K$;
upward triangles, $T=300 \, \rm K$; diamonds, $T=320 \, \rm K$;
downward triangles, $T=350 \, \rm K$; pentagons, $T=400 \, \rm K$.
Markers are colored according to the pressure of the liquid. For $T
\geq 300 \,\rm K$, increasing pressures lead to increasing values of
$-s_{2}^{\rm (tr)}$; for $T < 300 \,\rm K$, increasing pressures
first produce an increase of both order parameters
(negative-pressure states), followed by their simultaneous decrease
that partially stops at the minimum of the ordering locus beyond
which the translational order parameter starts growing again. The
dotted line has vanishing intercept (not shown) and was traced as a
guide for the eye to follow the asymptotic behavior of the order
parameters at high temperatures and low pressures. The inset shows a
comparison between the results obtained at $T = 300 \, \rm K$ with
108 (triangles) and 512 particles (crosses).}
\end{figure}
We note that the pair entropy already represents more than $80\%$ of
the total configurational entropy of TIP4P water at the highest
temperature and largest pressure that we have investigated.
Moreover, this fraction increases with decreasing temperatures and
pressures until $-s_{2}$ eventually overcomes $-s_{\rm
ex}$~\cite{saija2003}. On the other hand, orientational correlations
account for the overwhelming contribution to the pair entropy. We
found that this contribution also increases, with decreasing
temperatures and pressures, from a percent value of $\sim 67\%$ for
$T=400 \, \rm K$ and $P=10 \, \rm kbar$ to $\sim 79\%$, about which
it appears to saturate at low enough temperatures and pressures.
%
%
\begin{figure}
\includegraphics[width=8cm,angle=0]{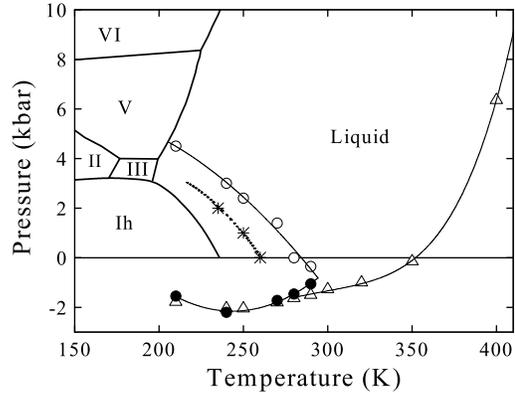}
\caption{\label{fig2} Phase diagram of the TIP4P model for water: solid circles and open triangles identify states
corresponding to the maxima observed in $-s_{2}^{\rm (tr)}(\rho)$ and $-s_{2}^{\rm (or)}(\rho)$, respectively; open
circles identify states corresponding to the minimum exhibited by $-s_{2}^{\rm (tr)}(\rho)$. Stars locate the states of
maximum liquid density. Lines traced across the data points are a guide for the eye. The solid-solid and solid-liquid
boundaries are sketched after the coexistence lines shown in Fig.~1 of~\cite{abascal2005} for the TIP4P model.}
\end{figure}
We now turn to a discussion of the thermodynamic behavior of $-s_{2}^{\rm (tr)}$ and $-s_{2}^{\rm (or)}$. The former
quantity behaves differently according to temperature. For $T \ge 300 \, \rm K$, the translational order parameter
monotonically increases - over the explored range - as a function of density or pressure, as is typically observed in an
ordinary simple fluid~\cite{errington2003}. Below ambient temperature, this behavior is still maintained at either low
or high enough pressures: for $T \lesssim 295 \, \rm K$, a window actually opens up within which $-s_{2}^{\rm (tr)}$
decreases with increasing pressures. As for the orientational order parameter, this quantity shows a maximum as a
function of density, which sharpens with decreasing temperatures. For $T \lesssim 280 \, \rm K$, the
maximum observed {\it in both quantities} falls - to within the numerical uncertainty of the calculations - around the
same density ($\sim 0.92 \, \rm g/cm^3$),

The nature of the correlation between translational and orientational order can be enlightened upon constructing a
two-parameter ordering map, as originally proposed by Torquato and coworkers in~\cite{torquato2000}. The map generated
by the order parameters introduced in this Letter is shown in Fig.~1, where we plotted $-s_{2}^{\rm (tr)}$ as a function
of $-s_{2}^{\rm (or)}$. We identify three distinct regimes according to the nature of the structural response of water
to compression. At high temperatures, both translational and orientational order are enhanced by pressure in low-density
water. However, at higher densities, a different structural condition develops, characterized by opposite behaviors of
the two order parameters: in this hybrid regime the translational order keeps increasing while the orientational order
decreases as the pressure is raised further and further. At low temperatures, the compression of the liquid may also
weaken both types of order: such a twofold anomalous behavior is observed in an intermediate range of densities whose
borders on the ordering locus coincide with either the relative or absolute maximum of $-s_{2}^{\rm (or)}$ or
$-s_{2}^{\rm (tr)}$, respectively, and with the minimum of $-s_{2}^{\rm (tr)}$. In passing, we note that the shape of
the ordering map is not appreciably modified by the size of the calculation (see the inset in Fig.~1).

The mapping of the $(T,P)$ domains corresponding to the different ordering regimes outlined above onto the phase diagram
of TIP4P water is shown in Fig.~2. The lower boundary of the structurally anomalous region is entirely located in the
negative pressure range. This threshold shifts to higher pressures with increasing temperatures. However, it is
interesting to note that this line traces, almost all over the range, a constant-density path ($\rho \simeq 0.92 \, \rm
g/cm^3$). We surmise that, in the light of the present results, this density can be interpreted as a sort of intrinsic
threshold - in the liquid phase - for the nucleation of a stable solid with perfect tetrahedral co-ordination.
Indeed, the density of the TIP4P Ih ice at ambient pressure ($0.937 \, \rm g/cm^3$)~\cite{abascal2005} is consistent
with this hypothesis. We also recall that the corresponding experimental value is $\sim 0.92 \, \rm g/cm^3$~\cite{petrenko}.

The upper boundary of the anomalous region shifts to lower pressures with increasing temperatures. Correspondingly, the
density decreases from $\sim 1.2 \, \rm g/cm^3$ to $\sim 0.95 \, \rm g/cm^3$. As seen from the figure, a
twofold anomalous structural behavior is observed in the liquid coexisting with ices Ih, III, and, partially, V. It is
rather natural to explain such an anomalous behavior as due to the presence of a persistent and extended hydrogen-bonded network.
Consequently, we are lead to identify the associated region as the LDW structural basin, as opposed to the HDW basin
where the translational and orientational order parameters show opposite trends as a function of the pressure. The
lower boundary of the HDW region, separating above ambient temperature thermodynamic regions where the angular
order is either enhanced or disrupted by pressure, crosses the $P=0$ axis at a temperature ($T\simeq 350 \, \rm K$)
that is close to the boiling point of TIP4P water at ambient pressure ($T\simeq 363 \, \rm K$)~\cite{vlot1999}. We also
note that the locus of density maxima lies entirely inside this region as also found by Errington and Debenedetti
through a tetrahedricity-based order representation~\cite{errington2001}.

On the basis of the analysis of the experimental data carried out by Soper and Ricci~\cite{soper2000}, one is lead to
identify the LDW-HDW crossover threshold with the state of the liquid whose oxygen-oxygen structure factor can be
represented by an equal-weight superposition of the ``asymptotic" LDW and HDW forms. At $T = 268 \, \rm K$, this
criterion yields a pressure of $\sim 1 \, \rm kbar$. The estimate obtained with the present method for the TIP4P model
at the same temperature is about $1.2 \, \rm kbar$. As for the thermodynamic slope of the LDW-HDW structural boundary,
our results are consistent with the numerical predictions of Saitta and Datchi~\cite{Saitta03}, who used a criterion
based on bond-angles distributions to trace the crossover line, and with the experimental findings of Li and
coworkers~\cite{li2005}.

In this Letter we have implemented, on a rigorous statistical-mechanical basis, an entropy-based theoretical framework
that allows a structurally unbiased description of ordering regimes in liquid water, in its thermodynamically stable
domain. The current individuation of LDW and HDW basins is in good agreement with neutron-diffraction results at $268
 \, \rm K$ and further clarifies other conflicting experimental data. We plan to extend this analysis to the study of
amorphous ices. However, the proposed metrics prospectively qualify the current method as a general tool for
investigating the nature of structural order in molecular liquids.

We thank Dr. J. Zielkiewicz for providing us with his numerical estimate of the orientational pair entropy for the TIP4P
model. One of the authors (R. E.) acknowledges the support of a Socrates/Erasmus grant and the scientific hospitality of
the Universit\'e Pierre et Marie Curie during the Erasmus stage.
%
%

\end{document}